\documentclass{sf2a-conf}
\usepackage{graphicx}
\begin{document}

\TitreGlobal{SF2A 2006}

\title{The Herschel galaxy reference survey}
\author{Boselli, A.}\address{Laboratoire d'Astrophysique de Marseille}
\author{the SPIRE extragalactic group}
%
\runningtitle{The Herschel galaxy reference survey}
\setcounter{page}{237}

\maketitle
\begin{abstract} 
In order to study the dust properties of different galaxies in the nearby Universe, the
SPIRE extragalactic group selected a volume limited (15$<$ $D$ $<$25 Mpc), complete 
sample of 313 galaxies spanning the whole range in morphological 
type (from ellipticals to late-type spirals)
and luminosity (8.5 $<$ log $L_H$ $<$ 12 L$_H\odot$, -22 $<$ $M_B$ $<$ -16) extracted from 2MASS, 
to be observed in guaranteed time with Herschel.  
The 250-360-520 $\mu$m SPIRE data, combined with those collected at other frequencies, will be used
to trace the dust properties of normal galaxies and provide a reference sample 
for studies at high redshift.
\end{abstract}
%
\section{Introduction}
During their evolution, the different stellar populations of galaxies produce and inject into the 
interstellar medium metals that congregate to form dust particles of different size and composition. 
This dust, heated by the general stellar radiation field, re-emit the absorbed energy in the far-IR.
In normal galaxies, those objects dominating in number the nearby universe, the emitting dust has a 
modified black body spectrum with a peak at $\sim$ 200 $\mu$m rapidly decreasing at longer wavelengths 
(Boselli et al. 2003). Because of the 
quiescent star formation activity of these objects, however, most of the dust has a relatively cold 
temperature ($\sim$  20 K) and is emitting in the submillimetric domain. The total dust amount of 
normal galaxies can thus be measured only by means of observations in the 200-1000 $\mu$m spectral range.\\
All physical (star formation activity, gas content, metallicity; Boselli et al. 2001; 2002; Zaritsky et al. 1994; 
Gavazzi et al. 2004), structural (concentration index, light profile; Boselli et al. 1997; Gavazzi et al. 2000; 
Scodeggio et al. 2002),
kinematical (shape of the rotation curve; Catinella et al. 2006) and stellar population 
(stellar spectral energy distribution; Gavazzi et al. 2002; Boselli et al. 2003)
properties of galaxies are strongly related to their total stellar mass (downsizing effect) and marginally
to their morphological type. Tracing the statistical dust properties of normal galaxies can thus be done 
by observing a large and complete sample of objects spanning the largest possible range in luminosity and Hubble
type.\\
Although accessible to ground based facilities such as SCUBA, the observation of such large 
samples of normal, quiescent galaxies would be too time consuming (more than 20 hours per galaxy) (Dunne et al.
2000). 
Given its large field of view (4'x8') and its sensitivity ($\sim$ 7 mJy for a 5 $\sigma$ detection for 
a point source in 1 hour integration), SPIRE on Herschel is the ideal instrument for
such a survey. For such a purpose, the SPIRE extragalactic group selected a volume limited, complete sample of
$\sim$ 300 galaxies to be observed in 100 hours of guaranteed time in the three SPIRE bands at 250, 360 and 520
$\mu$m.

\section{The sample and the observations}

To span the largest possible range in mass, the sample has been selected in the $K$ band (from 2MASS)
since near-IR luminosities are linearly related to the total dynamical mass of galaxies (Gavazzi et al. 1996). 
A distance constraint of 15 $<$ $D$ $<$ 25 Mpc has been introduced to construct a volume limited sample. This
distance range limits distance uncertainties due to local peculiar motions and secure the presence of
low-luminosity, dwarf galaxies, non accessible at higher redshift. To minimize 
galactic cirrus contamination, galaxies have been selected at high galactic latitude ($b$ $>$ $\pm$ 54$^o$)
and in low galactic extinction regions ($A_B$ $<$ 0.15; Schlegel et al. 1998).
Our primary sample contains all galaxies (from ellipticals to spirals) with $K$ $<$ 9 ($\sim$ 126 objects).
A secondary sample of $\sim$ 187 late-type galaxies with 9 $<$ K $<$ 12 has been added to cover the low
luminosity range. This secondary sample does not include the dust poor ellipticals whose emission would be hardly
detected within reasonable integration times.\\

\begin{figure}[ht]
\begin{center}
\includegraphics[width=9cm]{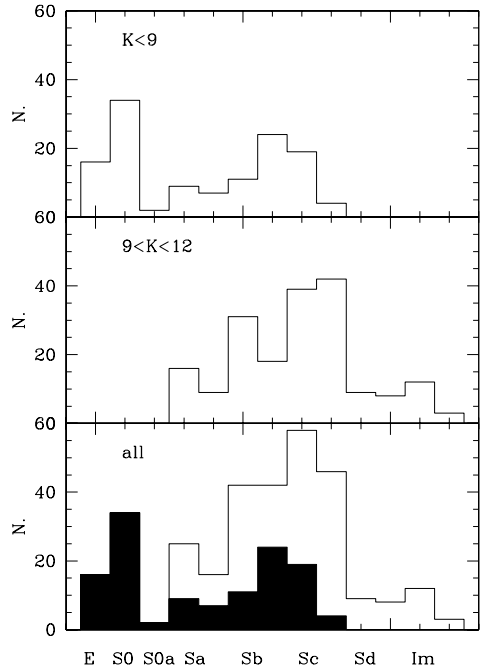}
\caption{The morphology distribution in the primary (upper panel), secondary (central panel) and total (lower
panel) Herschel galaxy reference sample.
}
\label{fig}
\end{center}
\end{figure}

Figure 1 shows the histogram of the morphology distribution of the selected sample. As selected, the sample
includes both relatively isolated and cluster object (Virgo and Fornax) and is thus ideal for 
a statistical study on the effects of the environment on the dust properties of galaxies (Figure 2). 
Dwarf irregulars and blue compact galaxies, here undersampled in spite of the selected large range 
in luminosity, will be the targets of another SPIRE key program (Madden et al., this conference).
  
\begin{figure}[ht]
\begin{center}
\includegraphics[width=9cm]{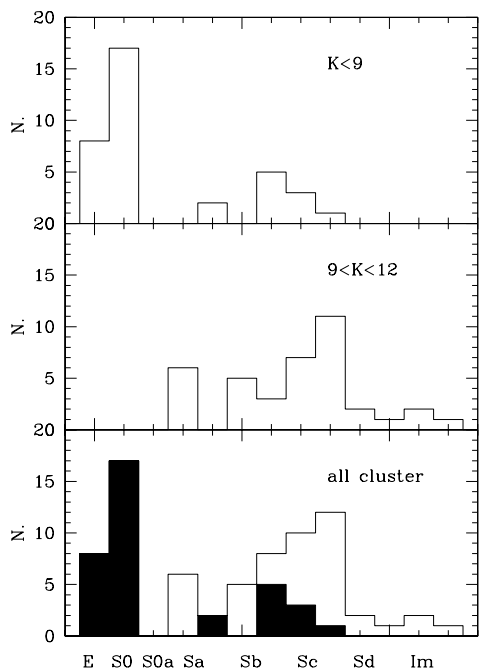}
\caption{The morphology distribution in the primary (upper panel), secondary (central panel) 
and total (lower panel) Herschel galaxy reference sample for Virgo and Fornax galaxies.
}
\label{fig}
\end{center}
\end{figure}

Pointed observations will be done with integration times of 30 minutes for ellipticals and lenticulars, 
and 12 minutes for late-types. Given a slewing overhead of 3 minutes per observation, this will ensure an
efficient use of the telescope. For early-type galaxies the integration time has been determined by assuming 
a lower limit to their total dust mass of $\sim$ 10$^4$ M$\odot$ as determined from IRAS observations 
(Bregman et al. 1998). With 30 minutes of integration a detection limit of 11 mJy at $\sim$ 4 $\sigma$ will be achived.
For spirals the integration time is dictated by the need of detecting the cold
dust emission outside the optical radius. By combining ISOPHOT (Alton et al. 1998) and SCUBA
(Valahakis et al. 2005) observations of extended sources 
with spectral energy distribution of normal galaxies of different type (Boselli et al. 2003), and assuming a
standard dust to gas ratio we estimate that 12 minutes of integration are sufficient to detect the cold dust
associated to the extended HI disc. With 12 minutes of integration time we will get to a detection limit of 
22 mJy at $\sim$ 5 $\sigma$.

\section{Science projects}

The selected sample will allow us to trace for the first time the variation of the cold dust properties 
(dust mass and temperature, dust to gas ratio,...) of normal galaxies along the Hubble sequence and as a function
of luminosity. Combined with multifrequency datasets, the SPIRE data will be used to study the role of dust in the
physics of the interstellar medium. Through the shielding of the interstellar radiation field, in particular
of the UV light, dust participate in the
cooling of the gas and thus plays a major role in the process of star formation. Dust is also an important 
catalyzer in the formation of the molecular hydrogen, and thus is critical entity for the study of the feedback.
Modeling the energetic balance between emitted and absorbed light, UV to sub-millimetric spectral energy
distributions of these 300 galaxies will be used to quantify the dust obscuration in different objects. 
This analysis will allow us to define standard recipes for correcting UV and optical data, a useful tool for
the interpretation of all modern surveys.\\ 
The presence of dust shells and discs in ellipticals will be used to study their hierarchical formation history.
By studying the relationship between dust mass and other global properties of ellipticals we will determine how
much dust is produced by the old stellar population and how much is the result of mergers.\\
The comparison of cluster and isolated galaxies will allow us to make a detailed study on the effects of the
environment on the dust properties of galaxies, and thus understand whether the hot and dense cluster
intergalactic medium can be polluted through the gas stripping process of late-type galaxies (Boselli \& Gavazzi
2006).
This dataset will be used to measure the local luminosity and dust-mass functions and distributions,
important benchmarks for the deep Herschel surveys planned by the SPIRE team, providing at the same time 
a unique reference sample for any statistical study.

\section{Corollary data}

The proposed analysis can be done only once corollary data covering the whole electromagnetic spectrum will be
available. Given its definition, the Herschel reference sample is easily accessible for ground based and
space facilities: the selected galaxies are in fact relatively bright ($m_B$ $<$ 15 mag) and extended 
($\sim$ 2-3 arcmin). UV data at 1500 and 2300 \AA ~will be taken from the GALEX all sky survey or the UV atlas 
of nearby galaxies (Gil de Paz et al. 2006), optical and near-IR
images will be secured by the SDSS (Abazajian et al. 2005) and 2MASS (Jarrett et al. 2003) 
surveys. These data will be used to trace the underlying stellar
population and reconstruct the star formation history of the target galaxies.\\
Far-IR data in the 10-200 $\mu$m spectral range, needed for
characterizing the whole dust emission and thus constraining dust masses and temperatures, 
will be taken from the AKARI/ASTRO-F survey at a similar spatial
resolution ($\sim$  20-30 arcsec, depending on the wavelength). Radio continuum data are already available from
the NVSS survey (Condon et al. 1998), while HI data will be obtained thanks to the ALFALFA survey (Giovanelli et al. 2005).\\
Dedicated spectroscopic observations in drift scan mode, needed to measure the metal content of the target
galaxies as well as their underlying stellar population at higher spectral resolution than with imaging,  
is under way with CARELEC at the OHP 1.93m telescope. We also started H$\alpha$ imaging observations
(needed to measure the present day star formation activity) at the San Pedro Martir 
Observatory.\\
Follow up X-ray (Chandra, XMM), CO (FCRAO, JCMT) and radio (Westerbork, VLA) observations of the target galaxies
are also in program.


\begin{thebibliography}{}
\bibitem{}Abazajian, K., et al$.$, 2005, 129, 1755
\bibitem{}Alton, P., Trewhella, M., Davies, J., et al., 1998, A\&A, 335, 807
\bibitem{}Boselli, A., Tuffs, R., Gavazzi, G., Hippelein, H. \& Pierini, D., 1997, A\&AS, 121, 507
\bibitem{}Boselli, A., Gavazzi, G., Donas, J. Scodeggio, M., 2001, AJ, 121, 753
\bibitem{}Boselli, A., Lequeux, J., Gavazzi, G., 2002, A\&A, 384, 33
\bibitem{}Boselli, A., Gavazzi, G., Sanvito, G., 2003, A\&A, 402, 37
\bibitem{}Boselli, A., Gavazzi, G., 2006, PASP, 118, 517
\bibitem{}Bregman, J., Snider, B., Grego, L., Cox, C., 1998, ApJ, 499, 670
\bibitem{}Catinella, B., Giovanelli, R., Haynes, M., 2006, ApJ, 640, 751
\bibitem{}Condon, J., et al., 1998, AJ, 115, 1693
\bibitem{}Dunne, L., Eales, S., Edmunds, M., Ivison, R., Alexander, P., Clements, D., 2000, MNRAS, 315, 115
\bibitem{}Gavazzi, G., Pierini, D., Boselli, A., 1996, A\&A, 312, 397
\bibitem{}Gavazzi, G., Franzetti, P., Scodeggio, M., Boselli, A., Pierini, D., 2000, A\&A, 361, 863
\bibitem{}Gavazzi, G., Bonfanti, C., Sanvito, G., Boselli, A., Scodeggio, M., 2002, ApJ, 576, 135
\bibitem{}Gavazzi, G., Zaccardo, A., Sanvito, G., Boselli, A., Bonfanti, C., 2004, A\&A, 417, 499
\bibitem{}Gil de Paz, A., Boissier, S., Madore, B., et al., 2006, ApJS, in press (astroph/0606440)
\bibitem{}Giovanelli, R., Haynes, M., Brian, K., et al., 2005, AJ, 130, 2598
\bibitem{}Jarrett, T., Chester, T., Cutri, R., Schneider, S. \& Huchra, J., 2003, AJ, 125, 525
\bibitem{}Schlegel, F., Finkbeiner, D., Davis M., 1998, ApJ, 500, 525  
\bibitem{}Scodeggio, M., Gavazzi, G., Franzetti, P., Boselli, A., Zibetti, S., Pierini, D., 2002, A\&A, 384, 812
\bibitem{}Valahakis, C., Dunne, L., Eales, S., 2005, MNRAS, 364, 1253
\bibitem{}Zaritsky, D., Kennicutt, R., Huchra, J., 1994, ApJ, 420, 87
\end{thebibliography}
\end{document}